# AGGREGATE COMBAT MODELING USING HIGH-RESOLUTION SIMULATION: THE "MEETING ENGAGEMENT" SCENARIO AS A CASE STUDY

**Sumanta K. Das, Pankaj Sati, and Rajiv Gupta[1]**

**Abstract. This paper illustrates a methodology for developing aggregated combat models using high-resolution simulation and the Markovian Lanchester process. The details of mathematical models involved in a high-resolution simulation process of a "Meeting Engagement" tactical scenario are presented, along with theoretical discussion on the Markovian Lanchester model. The output from the discrete event simulation model is used to estimate the attrition rates for an aggregated Lanchester model. A comparative study of various statistical estimation methods suitable for such estimation is also presented.**

## INTRODUCTION

Defence planners and decision makers use mathematical models to predict likely outcomes of combat dynamics. These mathematical models are generally represented in the form of a system of deterministic differential equations, which represent the gradual interaction and attrition process of the two sides. Lanchester in 1914 first introduced the concept of combat modeling using differential equations [1]. Many analysts have subsequently modified his original work to represent combat dynamics in modern warfare. Weiss [2] modified Lanchester's original work for aimed fire (such as by armour). Brackney [3] introduced the concept of area fire (such as by artillery). Helmold [4] has given a general form for homogeneous-force attrition rates (square / linear / logarithmic) and proposed a modification of Lanchester equations for modern warfare to account for inefficiencies of scale for the larger force when force sizes are grossly unequal.

Obtaining numerical values of attrition-rate coefficients (the rate at which an individual weapon-system type kills enemy targets of a particular type) is a major problem for applying the Lanchester model in practice. Two approaches have been originated in this respect [5]:

- use of analytical sub models, of the attrition process to compute the desired numerical values; and
- a statistical estimate, based on 'combat' data generated by a detailed combat simulation.

In reality, actual historical combat data is not easily available. Therefore, the practice is to use data generated either by combat field experiments or by a high-resolution combat simulation. In the latter approach, one uses combat data to compute statistical estimates of the attrition rate coefficients.

There are four principal statistical methods for computing such point estimates [6]: (a) method of moments estimation (MME) (b) maximum likelihood estimation (MLE) (c) Bayes estimation (BE) and (d) least square estimation (LSE). Of these four methods, only maximum likelihood estimation method has been used extensively for estimating attrition rate coefficients from combat simulation [7–11]. Since the original work of Clark [7], no significant theoretical improvement in the combat simulation approach has appeared in the open literatures. Clark in his work had assumed that every target type on a side had the same availability, for statistical estimation of model parameters, using time gap between two successive casualties, from the simulation model. There are no alternatives to such assumptions [12–13]. However, Taylor [11] has shown how to estimate attrition rate coefficients, without assuming that all target types on a side have the same target availability.

Comparison of high and low-resolution models and the need of aggregation are elaborated in the literature [14]. High-resolution modelling involves detailed design, high-resolution knowledge usage, narrow in-depth analysis for accuracy, reasoning and comprehension at a more atomic level, and simulating reality. Contrary to this, low-resolution modelling involves simplistic design, low-resolution knowledge usage, responding to mainly high-level questions, reasoning and comprehension with high-level variables, and abstracting "big picture". High-resolution modelling can also be used for informing, calibrating, or explaining low-resolution work.

The taxonomy of models, in particular high-resolution and low-resolution models, is widely discussed in the literature [15]. The focus of these works is on the connection between the strategic planning with detailed analysis. Aggregation and disaggregation are techniques that facilitate interactions at the same level of interactions. The paper [16] illustrates common approach used in aggregation using Lanchester theory as a basis. Requirements for theoretically consistent aggregation, disaggregation, and partial aggregation have also been described. Consistency of aggregation and disaggregation in models of combat is a very desirable property.

The objective of this paper is to estimate the attrition rate coefficients from simulation approach using various statistical estimators to compare their efficiencies (that is, with minimal variance as far as possible); considering a typical scenario such as a "Meeting Engagement" of armour combat.

The simulation approach has also been known as the hierarchy-of-models approach or fitted-parameter-analytical-model approach, such as the attrition-calibration (ATCAL) methodology implemented in the U.S.A Army's CEM model [13,17]. The simulation approach for estimating attrition-rate coefficients in Lanchester-type models (described graphically in Figure 1[11]) consists of the following:

- collecting information through multiple run of Monte-Carlo simulation model,
- set of aggregated Lanchester equations,
- statistical estimation of attrition-rate coefficients,
- situation matching/extrapolation methodology, and
- solution of aggregated Lanchester equations.

---

1

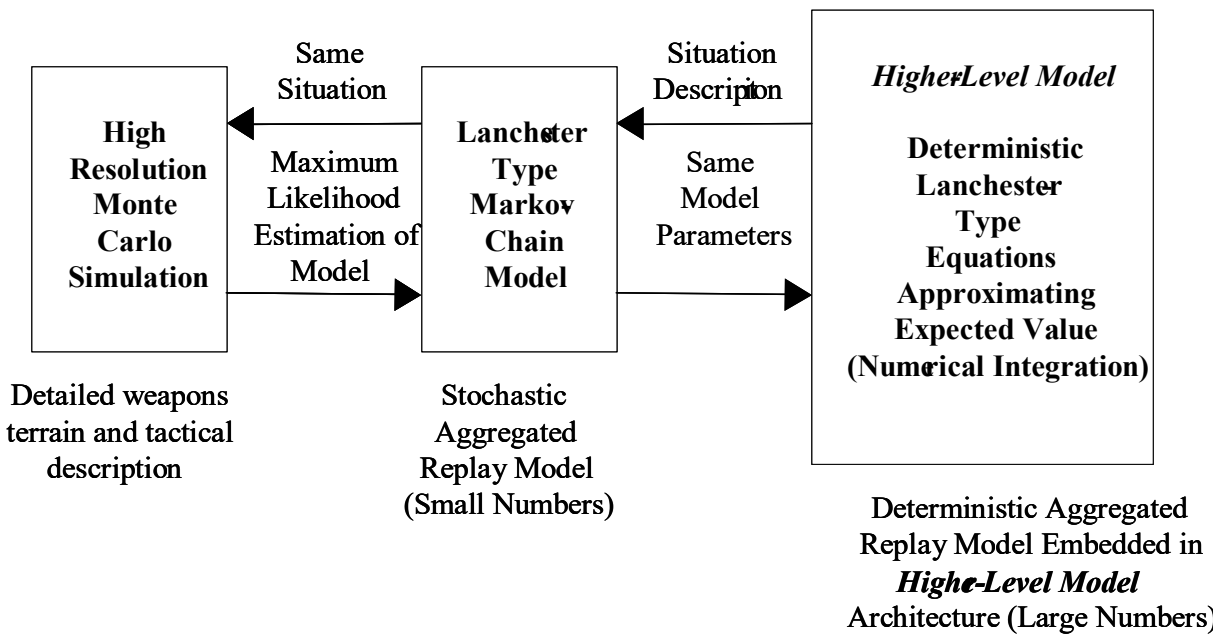

**Figure 1. Hierarchy-of-modelling concept.**

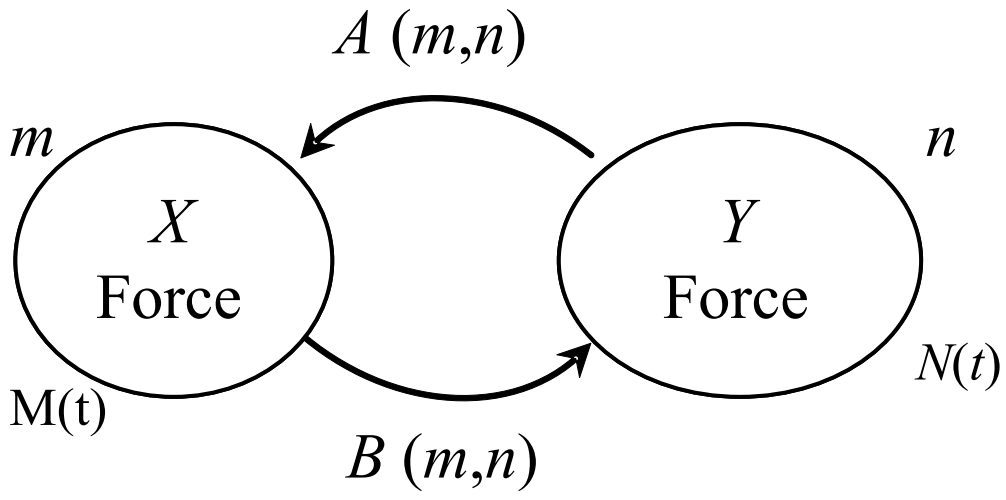

**Figure 2. Schematic of combat interactions for stochastic battle corresponding to the deterministic Lanchester type equations (1).**

## THEORY

For the case of two homogeneous forces, the following Lanchester-type equations are generally used in the high-resolution simulation [12, 18, 19] as given in (1).

$$\begin{cases} \dfrac{dx}{dt} = -A(x,y) \quad \text{with } x(0) = x_0 \\ \dfrac{dy}{dt} = -B(x,y) \quad \text{with } y(0) = y_0 \end{cases} \tag{1}$$

where $x(t)$ and $y(t)$ denote, respectively, the $X$ and $Y$ force levels at time t. Let us assume that there is no replacement or withdrawals and $A$ and $B$ are the attrition rates of the $X$ and $Y$ forces, respectively.

If we want to statistically estimate the attrition rates in the model (1) from simulation output data we must consider a stochastic version of the model in which casualties occur randomly over time.

Letting $M(t)$, a random variable, denote the integral number of $X$ combatants alive at time $t$ (with corresponding realization denoted as m) and similarly for $N(t)$ which pertains to the $Y$ force as shown in the Figure 2, we then have the following Kolmogorov equations for the evaluation of the state probability for $0 \le m \le m_0$ and $0 \le n \le n_0$ as given in Equation (2) [5].

$$\frac{d}{dt}P(t,m,n) = A(m+1,n)P(t,m+1,n) + B(m,n+1)P(t,m,n+1) - \{A(m,n) + B(m,n)\}P(t,m,n) \tag{2}$$

where $P(t,m,n)=P[M(t)=m,\ N(t)=n \mid M(0)=m_0,\ N(0)=n_0]$

Now, suppose attrition rates appear linearly in $A(m,n)$ and $B(m,n)$, as given in (3).

$$A(m,n)=ag_a(m,n) \text{ and } B(m,n)=bg_b(m,n) \tag{3}$$

where $g_a(m,n)=\{1-(1-P_A)^m\}n$ and $g_b(m,n)=\{1-(1-P_B)^n\}m$; $P_A$ and $P_B$ refer to the probability of target availability of $X$ and $Y$ respectively; and $a$ and $b$ are Lanchester attrition-rate coefficients and are constants.

## STATISTICAL ESTIMATION OF ATTRITION RATE

Consider now that we have run a simulation for L number of times and have recorded the times at which casualties have occurred (and also the type of each casualty). Let us run this stochastic simulation until a total of K casualty has occurred. The total time that the simulation will have been run is a random variable. Let us also denote (for $k_L$=1,2…$K$) the time (a random variable) at which the $k$th casualty occurs as $T_{k_L}$ (with realization $t_{k_L}$).

For the case of two homogeneous forces, the four parameters $a$, $b$, $P_A$, and $P_B$ for the model given by (3) are to be estimated from the output of the high- resolution combat simulation. One can develop statistical estimates such as MME, MLE, BE, LSE for these parameters. If the times at which each casualty has occurred (for both sides) have been recorded, then one can develop statistical estimates for these four model parameters. Let $C^X_{K_L}$, $C^Y_{K_L}$ denote two indicator variables at time $t_{k_L}$ of $X$ and $Y$ casualty respectively as defined in the (4).

$$\begin{aligned} C^X_{K_L} &= \begin{cases} 1, & \text{if } k^{\text{th}} \text{ casualty is an } X \text{ combatant} \\ 0 & \text{otherwise} \end{cases} \\ C^Y_{K_L} &= \begin{cases} 1, & \text{if } k^{\text{th}} \text{ casualty is an } Y \text{ combatant} \\ 0 & \text{otherwise} \end{cases} \end{aligned} \tag{4}$$

then $C^X_{T_L} = \sum_{k=1}^{K} C^X_{k_L}$ ; $C^Y_{T_L} = \sum_{k=1}^{K} C^Y_{k_L}$ ; and $C^X_{T_L} + C^Y_{T_L} = K$ .

Let $m_{k_L}$ is the realization of the number $X$ combatants just after the occurrence of the $k$th casualty in the $L^{\text{th}}$ simulation run. Similarly $n(t_{k_L})$ is the realization of $n_{k_L}$, In other words, there are $m_{k_L}$ $X$ combatants and $n_{k_L}$ $Y$ combatants "alive" during the interval $[t_{k_L}\ t_{k_{L+1}})$ for $k_L$=0,1,…, $K$–1.

Using the data $t_{1_L}, \ldots, t_{k_L}, C^X_{1_L}, \ldots, C^X_{K_L}\ C^Y_{1_L}, \ldots, C^Y_{K_L}$ we will now develop statistical estimates of $a$ and $b$ denoted by different methods of statistical estimations.

### Method of Moments Estimation (MME)

This is the simplest method. The method leads to estimates that are quickly and easily computed. MME is very straightforward and can be easily implemented in software. This is a technique for constructing estimators of the parameters that is based on matching the sample moments with the corresponding distribution moments. The disadvantage of this method is that MMEs are not necessarily sufficient statistics—that is, they sometimes fail to take into account all relevant information in the sample.

Let times between two casualties is exponentially distributed. From the stochastic model given in (2), the times between two successive *X* casualties or time to kill a *X* combatant by *Y* force is $\frac{1}{A(m_{k_L},n_{k_L})}$. With the present simulation set up, $C^X_{K_L}$ casualties will take time that is equal to $\frac{C^X_{K_L}}{A(m_{k_L},n_{k_L})}$; which is also equal to $t_{k_L}$. Hence the first-order moment of the simulated data is MME estimator of the attrition rate coefficients, which is derived as in (5).

$$\hat{a}_{MME} = \frac{1}{L}\sum_{1}^{L}\frac{C^X_{K_L}}{t_{k_L}(1-(1\text{-}\hat{P}_A)^{m_{k_L}})n_{k_L}} \quad (5)$$

**Maximum Likelihood Estimation (MLE)**

The MLE estimation begins with writing a mathematical expression known as the likelihood function of the sample data. Loosely speaking, the likelihood of a set of data is the probability of obtaining that particular set of data, given the chosen probability distribution. This expression contains the unknown parameters. The values of these parameters that maximize the sample likelihood are known as the "maximum likelihood estimates" or MLE.

We assume that casualty process can be considered as two Poisson processes. Let *S* denote the two-consecutive casualties, then the probability density function or pdf of the time to an *X* or *Y* casualty is given by (6).

$$f_S(s) = \{A(m,n)+B(m,n)\}e^{-(A(m,n)+B(m,n))s} \quad (6)$$

To construct the likelihood function, we observe that casualties have occurred at times $t_{1_L}$, … , $t_{k_L}$, there being a total of $C^X_{K_L}$, *X* casualties and $C^Y_{K_L}$, *Y* casualties at $L^{th}$ simulation. Occurrence of the $k^{th}$ casualty (either *X* or *Y*), which represents a transition from battle space ($m_{k_L}-1, n_{k_L}-1$) to ($m_{k_L}, n_{k_L}$) will contribute to the likelihood function as derived in (7).

$$\begin{aligned}L(a,b) = \prod_{k=1}^{K}\left(A(m_{k_{L-1}},n_{k_{L-1}})\right)^{C^X_{k_L}}\left(B(m_{k_{L-1}},n_{k_{L-1}})\right)^{C^Y_{k_L}} \times \\ \exp[-(A(m_{k_{L-1}},n_{k_{L-1}})+B(m_{k_{L-1}},n_{k_{L-1}}))\times \\ \{t_{k_L}-t_{k_{L-1}}\}\end{aligned} \quad (7)$$

Thus, computing the natural logarithm of the likelihood function and setting its first derivative with respect to *a* and *b* equal to zero, one finds that the MLE for *a* is derived as in (8).

$$\hat{a}_{MLE} = \frac{c^X_{T_L}}{\sum_{k=1}^{K}\left\{1-\left(1-\hat{A}\right)^{m_{k_{L-1}}}\right\}n_{k_{L-1}}(t_{k_L}-t_{k_{L-1}})} \quad (8)$$

MLE has very desirable large sample properties. They become minimum variance estimators as the sample size increases also they follow normal distribution. The limitation of MLE is that specialized software is required for solving complex non-linear solution.

**Bayesian Estimation (BE)**

Bayes estimator or decision rule maximizes the a posteriori expected value of a utility function or minimizes the a posteriori expected value of a loss function.

Let $\pi(\boldsymbol{\theta})$ is the prior distribution of attrition rate coefficients, where $\boldsymbol{\theta}$ is the parameter space, $f(t_{k_L}|\boldsymbol{\theta})$ is the density function of $t_{k_L}$ given $\boldsymbol{\theta}$, where *t* is a random variable represents the time at which a casualty has occurred in $L^{th}$ simulation run. Then the joint distribution of time and attrition rate coefficients is derived as in (9).

$$u(t_{1_L},t_{2_L},...,t_{k_L},\boldsymbol{\theta}) = \pi(\boldsymbol{\theta})\prod_{k=1}^{K}f(t_{k_L}\,|\,\boldsymbol{\theta}) \quad (9)$$

The marginal distribution of $t_{1_L}$, … , $t_{k_L}$ is derived as in (10).

$$g(t_{1_L},t_{2_L},...,t_{k_L}) = \int_{-\infty}^{\infty}\pi(\boldsymbol{\theta})\prod_{k=1}^{K}f(t_{k_L}\,|\,\boldsymbol{\theta}) \quad (10)$$

The a posteriori distribution of $\boldsymbol{\theta}$ given the $t_{k_L}$'s is derived as in (11).

$$h(\boldsymbol{\theta}\,|\,t_{1_L},t_{2_L},...,t_{k_L}) = \frac{u(t_{1_L},t_{2_L},...,t_{k_L},\boldsymbol{\theta})}{g(t_{1_L},t_{2_L},...,t_{k_L})} \quad (11)$$

Rewriting the equation $\prod_{k=1}^{K}f(t_{k_L}\,|\,\boldsymbol{\theta})$ = *L*(*a*,*b*); as the likelihood function of the $t_{k_L}$'s, we have the a posteriori probability of attrition rate coefficients are derived as in (12).

$$h(\boldsymbol{\theta}\,|\,t_{1_L},t_{2_L},...,t_{k_L}) = \frac{\pi(\boldsymbol{\theta})\prod_{k=1}^{K}f(t_{k_L}\,|\,\boldsymbol{\theta})}{\int_{-\infty}^{\infty}\pi(\boldsymbol{\theta})\prod_{k=1}^{K}f(t_{k_L}\,|\,\boldsymbol{\theta})} \quad (12)$$

Assuming uniform priori distribution with parameter 0 and 1, of attrition rate coefficients the Bayes estimates of the attrition rate coefficients are derived as in (13).

$$\hat{a} = \int_0^1 a h(\boldsymbol{\theta}\,|\,t_{k_L})d\boldsymbol{\theta} = \frac{\frac{1}{(g_a(m_{k_L},n_{k_L})t_{k_L})^{C^X_{T_L}+2}}\Gamma(C^X_{T_L}+2)}{\frac{1}{(g_a(m_{k_L},n_{k_L})t_{k_L})^{C^X_{T_L}+1}}\Gamma(C^X_{T_L}+1)} \quad (13)$$

where $\Gamma(\,)$, is the Gamma function and $\Gamma(\alpha+1)=\alpha\Gamma(\alpha)$.

Hence $\hat{a} = \frac{C^X_{T_L}+1}{\left\{1\text{-}\left(1\text{-}\hat{P}_A\right)^{m_{k_L}}\right\}n_{k_L}t_{k_L}}$ the average of *L* simulations gives the Bayes estimate of *a* as:

$$\hat{a}_{BAYES} = \frac{1}{L}\sum_{1}^{L} \frac{C_{T_L}^{X}+1}{\left\{1-\left(1-P_A\right)^{m_{k_L}}\right\} n_{k_L} t_{k_L}} \tag{14}$$

**Least Square Estimation (LSE)**

Least square estimation (LSE) can be applied more generally than MLE or BE. The disadvantages of this estimation are: it is not applicable to censored data, it has less desirable optimality properties than MLE and BE, and it is sensitive to starting value. The advantage of this estimation is that it is easily implemented in software.

The Gauss-Markov form of (1) is:

$$\frac{dx}{dt} = -A(x,y) + e \tag{15}$$

which is equivalent to the common matrix form in (16).

$$\mathbf{X} = \mathbf{Y\theta} + \mathbf{e} \tag{16}$$

where: $\mathbf{X} = (m_{1_L}, m_{2_L}, m_{k_L})^T$,

$\mathbf{Y} = (g_a(m_{1_L}, n_{1_L}), g_a(m_{2_L}, n_{2_L}), g_a(m_{k_L}, n_{k_L}))^T$

$\mathbf{e}$ ~ **N**ormal $(\mathbf{0},\mathbf{I})$ and $\mathbf{\theta} = (a,b)^T$. Let $\mathbf{\theta}^*$ denote any arbitrary vector. This serves to define a vector of errors, or residuals as derived in (17).

$$\mathbf{e}^* = \mathbf{X} - \mathbf{Y\theta}^* \tag{17}$$

The least-square principle for choosing $\mathbf{\theta}^*$ is to minimize the sum of the squared residuals $\mathbf{e}^{*T}\mathbf{e}^*$.

$$\mathbf{e}^{*T}\mathbf{e}^* = (\mathbf{X} - \mathbf{Y\theta}^*)^T(\mathbf{X} - \mathbf{Y\theta}^*)$$

$$= \mathbf{X}^T\mathbf{X} - \mathbf{2\theta}^{*T}\mathbf{Y}^T\mathbf{X} + \mathbf{\theta}^{*T}\mathbf{Y}^T\mathbf{Y\theta}^{*T}$$

thus, $\dfrac{\mathbf{d(e^{*T}e^*)}}{\mathbf{db}^*} = -\mathbf{2Y}^T\mathbf{X} + \mathbf{2Y}^T\mathbf{Y\theta}^*$ make it equal to $\mathbf{0}$ vector we get the least square estimate of attrition rate coefficients as derived in (18).

$$\hat{\mathbf{a}}_{\mathbf{LS}} = (\mathbf{Y}^T\mathbf{Y})^{-1}\mathbf{Y}^T\mathbf{X} \tag{18}$$

For estimating target availability ($P_A$), we considered a continuous Markov Chain line-of-sight (LOS) process. A Markovian chain LOS model is illustrated in Figure 3. Three states of targets are defined in this Markovian chain LOS model. These are Visible and Acquired (VA), Visible and Not-Acquired (VNA), and Invisible (I) state. Transition from one state to another is possible. Only a two-way transition exists between VNA and I states. The transition rates ($\lambda$, $\eta$, $\mu$, $\iota$) between two states are shown in Figure 3. For the steady-state probability that the particular target is visible and acquired (that is, the target is available) is taken from [11,21] as in (19)

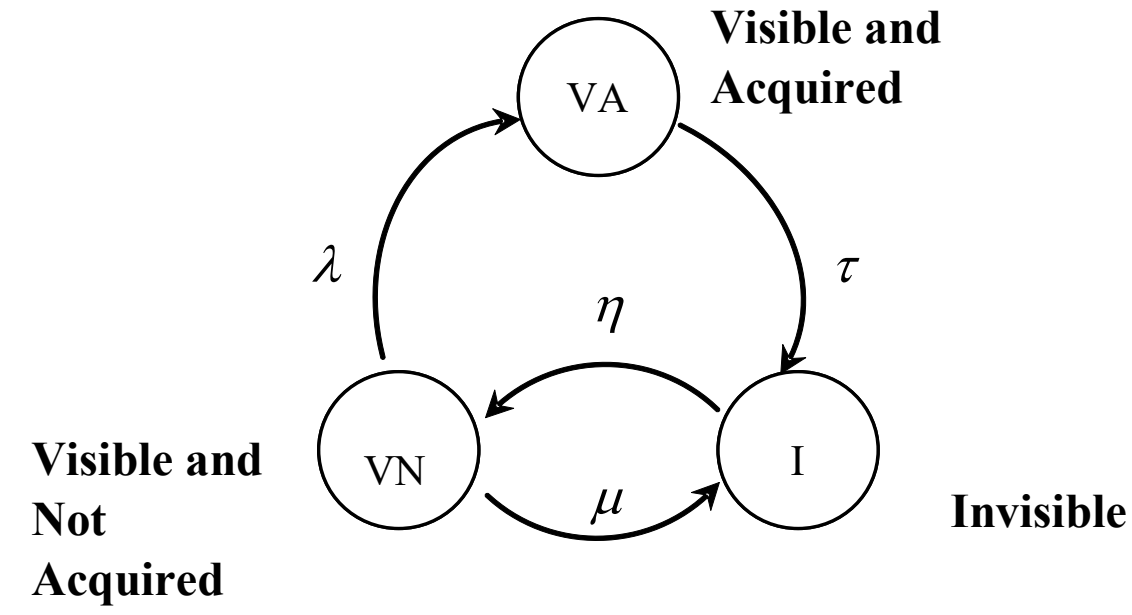

**Figure 3. Markov-chain model for interaction of LOS and target-acquisition processes.**

$$P_A = \frac{\eta_{X_iY_j}\lambda_{X_iY_j}}{\left(\eta_{X_iY_j} + \mu_{X_iY_j}\right)\left(\lambda_{X_iY_j} + \mu_{X_iY_j}\right)} \tag{19}$$

An estimate for target availability, for example, is derived as in (20).

$$\hat{P}_A = \frac{\hat{\eta}_{X_iY_j}\hat{\lambda}_{X_iY_j}}{\left(\hat{\eta}_{X_iY_j} + \hat{\mu}_{X_iY_j}\right)\left(\hat{\lambda}_{X_iY_j} + \hat{\mu}_{X_iY_j}\right)} \tag{20}$$

where the estimate for $\eta$ is given by the mean time that the target is in the invisible state, and similarly for $\mu$ and $\lambda$. length of time that a target was invisible. For a high-resolution simulation one has to consider transitions occurring directly from the invisible state to a target being acquired. A modification of the above Markov-chain model has been given in the references [11, 21]. Denoting the transition rate from the state of the target being invisible to it being visible and acquired as $\tau$, one finds that an estimate for target availability, for example, for the modified model is as in (21).

$$\hat{P}_A = \frac{\hat{\eta}_{X_iY_j}\hat{\lambda}_{X_iY_j} + \hat{\tau}_{X_iY_j}\left(\hat{\lambda}_{X_iY_j} + \hat{\mu}_{X_iY_j}\right)}{\left(\hat{\tau}_{X_iY_j} + \hat{\eta}_{X_iY_j} + \hat{\mu}_{X_iY_j}\right)\left(\hat{\lambda}_{X_iY_j} + \hat{\mu}_{X_iY_j}\right)} \tag{21}$$

Once target availabilities have been estimated, one can readily estimate the attrition rate coefficients.

## METHODOLOGY

Simulation is commonly used to study large, complex systems like combat. There are two different models for moving a system through time: the “fixed time-step” and the “event-to-event” model. Most of the simulations of ground combat use both the time-step method for the target-acquisition process and the event-step method for all other processes. The basic difference between Lanchester approach and simulation approach is about time. The mathematical approach has emphasized the use of continuous differential function. Simulation treats conflict as series of discrete events (discrete-time process).

Figure 4 illustrates the schematic diagram of the presented methodology. For the case of two homogeneous forces, in order to compute the estimates of attrition rate coefficients ($a$,$b$) one must first collect the causality data with simulation clock time from the output of the high-resolution simulation for both forces. For the case of heterogeneous forces, it is convenient to first compute the estimates for the Markov-chain model for the LOS and target-acquisition process for every firer-target pair from which the estimates for target availability can be computed. In the simplest case i.e. homogeneous forces, one must first collect the required data from the output of the high-resolution simulation in order to estimate the transition rates of this Markov chain as shown in Figure 3. From this figure, it is clear that timing of various events which are associated with transitions from one state to another namely the time at which a target in the invisible state becomes visible, the time at which a visible target becomes invisible, and the time at which a visible target becomes acquired are required to be collected. Data must be collected for every firer-target pair. Additionally, peculiarities of the high-resolution combat simulation required the collection of the length of time that a target was invisible for a target that was acquired in one sensor scan. The mean value of this quantity then estimated the rate $\tau$ at which targets formerly invisible were acquired. One must also collect the data about the simulation time, casualty, and shooter for all casualties.

The collected data is analysed to estimate Lanchester equations coefficients using different estimators as described in previous section. The best estimator is selected on the basis of their confidence level. Once parameters of Lanchester equations are chosen, aggregated models can be replayed.

## CASE STUDY

In order to present an example of the attrition rate estimation process, a high-resolution simulation of a combat scenario pertaining to armour battle, viz., “Meeting engagements” is chosen, in which the attacker has one squadron of tanks moving towards the objective. The defender has a half-squadron of tanks, a half-platoon of BMPs, and one platoon of infantry linearly deployed in the trenches with RCL. The combat forces meet in an open ground and the battle commences with both the armour advancing towards each other to inflict maximum casualties on each other. This is an integrated scenario in which effectiveness of both types of weapons, namely Tanks and BMPs, can be analyzed, under various conditions.

Various models (Detection, Line of sight, direct firing by Armour, and RCLs) run in the background and interact with each other to simulate the combat dynamics. The simulation is discrete-event fixed time-step. Entities in the simulation are soldiers and weapon systems. The events are detection, hit, and kill. Software is developed for simulating this scenario and GUI is shown in the Figure 5. Inputs to GUI are data/conditions pertaining to environment, terrain, and termination criteria of the simulation. The output are casualties in each simulation run, average casualties, percentage casualty, kill types of tanks (M Kill, F Kill, K Kill) and performance index (ratio of defender percentage casualties to attacker percentage casualties).

Detection model considers various parameters like target contrast, environment conditions, terrain, relative position of observer and target, and equipment characteristics to assess the probability of detection based on the underlying physics of the sensors. Necessary conditions for a direct fire engagement are: availability of line-of-sight, target to be within visibility range, and firing ranges. Coordination among tanks within a troop and coordination among troops are considered in the model.

High Resolution Simulation Models

Inputs

ORBAT

Weather

Scenario

Terrain Model | Mines

Armour | Artillery

RCL | Suppression

Detection/Acquisition

Simulation results

$\lambda, \eta, \mu, \tau, t_k$

$m_k, n_k$

Estimate a, b

Replay

Replay results

**Figure 4. Schematic diagram of the methodology.**

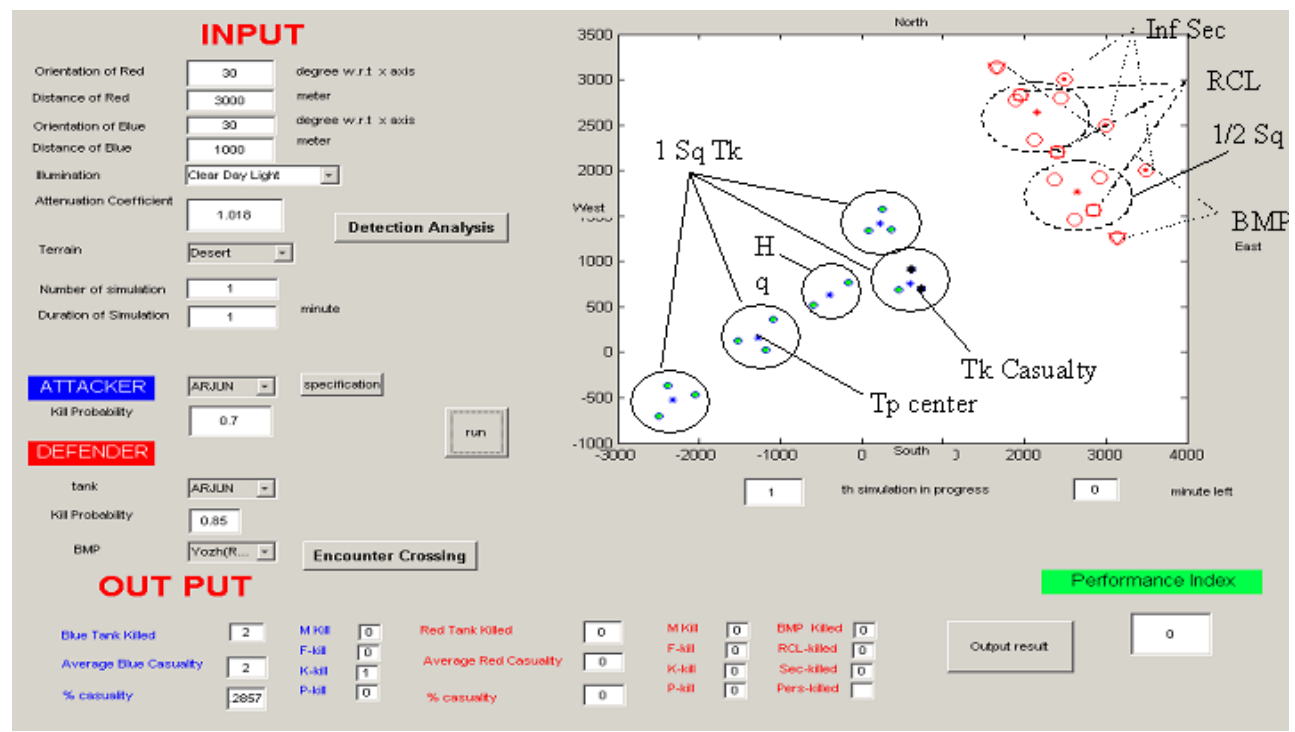

**Figure 5. Initial disposition of forces at start of battle considered in this work.**

Whether a tank will fire a shot on a target, is dependent on the target's availability or exposure time and target acquisition time required by the firer. These factors are sensitive to terrain, target and firer's status (static or moving) and other environmental conditions. This has been modelled accordingly. The number of shots to be fired at any instance on the target has been modelled considering its exposure time, acquisition time, flight time of shot, etc. Tank being a direct fire weapon, the hit probability of main gun depends on firing accuracy or aiming errors. The impact points of the shots are assumed to follow a circular normal distribution. Thus, the firing process has been modelled considering target projection based on its size, status (static or moving), and location (range and direction), using (22).

$$P_h = \iint_C p(x,y)\,dxdy = \frac{1}{2\pi\sigma^2}\iint_C \exp\left[-\frac{x^2+y^2}{2\sigma^2}\right]dxdy \quad (22)$$

where $P_h$ is hit probability, $p(x,y)$ joint distribution function of random shoot in horizontal ($x$) and vertical ($y$)directions, $c$ is the target area and $\sigma$ is the firing dispersion. For simplicity we assume that random shoot follows circular normal distribution.

Kill probability is related to single-shot kill probability (SSKP) of the firer and self-protection index (SPI) or survivability of the target, related to its armour. After generating the status of the event whether it is hit or missed, the casualty status of the event is generated. Based on this it is declared whether a tank is killed or alive using simulation.

This simulation is implemented in MATLAB ®[22]. Output of simulation data (simulation run, simulation clock time, casualties, remaining force strength) is stored in a database and tools are created to analyze the data, as described in the methodology. These tools are implemented in C language.

## RESULTS

The scenario described above was input into the system. Results have been generated based on 500 runs with the following scenario parameters: terrain (desert), visibility (800 m – 1,000 m), tank target size (height: 3 m, hull height: 1.47 m, width: 3.87 m, track width: 1.27 m, hull length: 6.95 m), target engagement ranges (600 m – 1,500 m), detection probabilities (0.5–0.75) and kill probabilities (0.8-1.0). The data addresses most of the variance in the scenario with an allowance of randomness on confirmation of events: detection, hit, and kill. This randomness is stabilised with 500 repetitions.

Numerical results for armour combat simulation, as described in the case study are shown in Figure 6. Force levels computed according to the aggregated model given by (1), denoted as Lanchester attrition rates are also shown in the figure. Similar results have been generated for other entities also, namely BMPs, RCLs and infantry units. This result is a realization of force levels for a particular simulation run. In the figure, the legend on the right-hand side describes the weapon systems involved in the scenario. For each unit, the first letter "S" refers to data relating to simulation run and "L" for the Lanchester model.

The statistical estimates of attrition rate coefficients for different methods obtained from the simulation results are given in Tables 1 and 2, with their standard error (SE) and 95% confidence level, for attacker and defender, respectively. By analyzing Tables 1 and 2, we conclude that the previous assumption that MLE always gives better results is not correct. From both tables, the LSE estimator performs better. Although the performance of the MME method is good, this estimation may sometimes give biased results. The BE method does not give any improvement in the present situation.

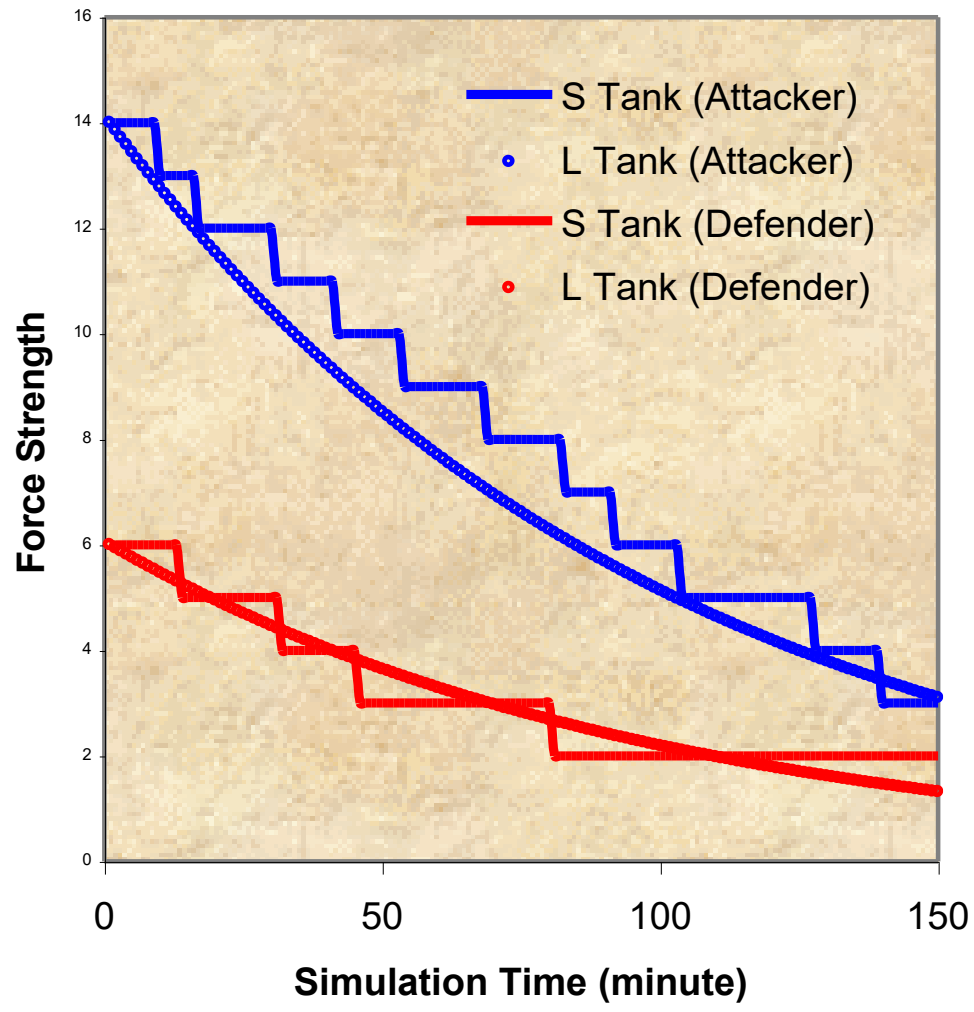

**Figure 6. Force-level decays for aggregated Lanchester equations (1) compared with realization of original simulation battle.**

| Estimator | Attrition Rate | S.E. | Confidence Level (95%) |
|---|---|---|---|
| MME | 0.0168 | 0.0086 | 0.00151 |
| MLE | 0.0127 | 0.0215 | 0.00377 |
| BE | 0.0197 | 0.0105 | 0.00185 |
| LSE | 0.0117 | 0.0002 | 0.00004 |

**Table 1. Statistical estimates of attrition rate coefficients of the attacker.**

| Estimator | Attrition Rate | S.E. | Confidence Level (95%) |
|---|---|---|---|
| MME | 0.00552 | 0.00370 | 0.00064 |
| MLE | 0.00308 | 0.00115 | 0.00020 |
| BE | 0.00694 | 0.00424 | 0.00074 |
| LSE | 0.00202 | 0.00008 | 0.00001 |

**Table 2. Statistical estimates of attrition rate coefficients of the defender.**

| Estimator | Attrition Rate | SE | Confidence Level (95%) |
|---|---|---|---|
| MME | 1.86193 | 0.6509 | 0.1050 |
| MLE | 1.85185 | 0.6480 | 0.1045 |
| BE | 1.86775 | 0.6526 | 0.1053 |
| LSE | 1.84749 | 0.6467 | 0.1043 |

**Table 3. Descriptive statistics of difference of simulated and aggregated model results of the defender.**

| Estimator | Attrition Rate | SE | Confidence Level (95%) |
|---|---|---|---|
| MME | 2.2698 | 1.8926 | 0.3054 |
| MLE | 2.2628 | 1.8905 | 0.3050 |
| BE | 2.2749 | 1.8941 | 0.3056 |
| LSE | 2.2610 | 1.8900 | 0.3049 |

**Table 4. Descriptive statistics of difference of simulated and aggregated model results of the attacker.**

| | Simulated Casualty | Lanchester Casualty |
|---|---|---|
| Mean Casualty | 8.06( $\bar{x}$ ) | 7.23( $\bar{y}$ ) |
| Variance of Casualty | 10.90 ( $s_1^2$ ) | 9.76( $s_2^2$ ) |
| Observations | 150 ($m$) | 150 ($n$) |
| $H_o$: Mean Difference | 0 ($d$) | |
| df | 297 | |
| $t$ Stat | 2.22 | |
| $P$ ($T$<=$t$) one-tail | 0.013 | >0.01; $H_o$ Accepted |
| $t$ Critical one-tail ($\alpha$=0.01) | 2.34 | >$t$ stat; $H_o$ Accepted |

**Table 5. Test of significance for comparing simulated casualty and Lanchester casualty for the attacker (blue).**

| | Simulated Casualty | Lanchester Casualty |
|---|---|---|
| Mean Casualty | 3.13($\bar{x}$) | 3.09($\bar{y}$) |
| Variance of Casualty | 1.81 ($s_1^2$) | 1.80($s_2^2$) |
| Observations | 150 ($m$) | 150 ($n$) |
| $H_o$: Mean Difference | 0 ($d$) | |
| df | 298 | |
| $t$ Stat | 0.227 | |
| $P$ ($T$<=$t$) one-tail | 0.41 | >0.01; $H_o$ Accepted |
| $t$ Critical one-tail ($\alpha$=0.01) | 2.34 | >$t$ stat; $H_o$ Accepted |

**Table 5. Test of significance for comparing simulated casualty and Lanchester casualty for defender (red).**

For selecting the best statistical estimation method in this context, we have calculated the differences of simulation results with predicted results obtained through Lanchester aggregated model. Descriptive statistics of the difference for both defender and attacker for different methods are given in Tables 3 and 4. In both the cases, we find that least-square estimator provides a smaller confidence interval. So we suggest that for estimating attrition rate coefficients from high-resolution combat simulation data, the least-square estimate is better and easy to implement.

It can be observed from Figure 6 that simulation results depict a flat line with respect to residual force after a certain stage of conflict. It is due to combat termination criteria based on residual force ratios. Lanchester's model does not depict a similar trend due to continuous nature of the curve considered therein, irrespective of the force ratio. To check the statistical significance of the difference of simulation results with aggregated Lanchester model result, we measure the *t*-statistic assuming unequal variances of casualty pattern. The *t*-statistics for both the attacker (blue) and defender (red) are given in Tables 5 and 6. From both the tables it is clear that there is no significant difference between simulated and aggregated Lanchester results at 99% confidence level.

The following formula is used to determine the value of *t* test statistic:

$$t = \frac{\bar{x} - \bar{y} - d}{\sqrt{\frac{s_1^2}{m} + \frac{s_2^2}{n}}} \tag{23}$$

The following formula is used to approximate the degrees of freedom (df).

$$df = \frac{\left(\frac{s_1^2}{m} + \frac{s_2^2}{n}\right)^2}{\frac{(s_1^2/m)^2}{m-1} + \frac{(s_2^2/n)^2}{n-1}} \tag{24}$$

The result of the calculation is usually not an integer; we use the nearest integer to obtain a critical value from the *t* table.

From Table 5 we see that there is no significant difference between simulated casualty and Lanchester casualty of the attacking forces (blue) at 1% level of significance. Thus, we can say that the Lanchester model can produce statistically similar result as simulation model. Similarly, we calculate the underlying *t*-statistic between simulated casualty and Lanchester casualty of defender (red). This result is shown in Table 6. Examining Table 6, we conclude that there is no significant difference between simulated casualty and Lanchester casualty of the defender (red) at 1% level of significance.

## CONCLUSIONS

This paper presents the salient features of the high-resolution simulation approach for modelling aggregated combat dynamics. It presents a methodology for estimating attrition rates based on results from high-resolution simulation. It then presents a comparative study of different statistical estimation methods for attrition-rate coefficients using more-detailed LOS/acquisition data, extracted from the high-resolution combat simulation. After statistical analysis of the results, it has been concluded that the least-square estimate provides better results for aforesaid combat process. This work can be extended for building relationship between casualty rates and force ratios. Such relationships will help in modelling combat dynamics at different levels of force sizes (Regiment, Brigade, Division, and so on) and extrapolating the combat outcomes both upwards and downwards.